\documentclass[bibnotes,prb,amsmath,citeautoscript,twocolumn,
               preprintnumbers,amssymb,amsmath,showpacs,floatfix,
               letterpaper]{revtex4}
\usepackage{graphicx} 

\def\bra#1{\mathinner{\langle{#1}|}}
\def\ket#1{\mathinner{|{#1}\rangle}}

\begin{document}


\title{Non-$d^0$ Mn-driven ferroelectricity in antiferromagnetic BaMnO$_3$}
  \author{James M.\ Rondinelli}
     \email[Address correspondence to: ]{rondo@mrl.ucsb.edu}
  \affiliation{Materials Department, University of California, Santa Barbara, 
	       CA, 93106-5050, USA}
  \author{Aaron S.\ Eidelson}
  \affiliation{Materials Department, University of California, Santa Barbara, 
	       CA, 93106-5050, USA} 
  \author{Nicola A.\ Spaldin}
  \affiliation{Materials Department, University of California, Santa Barbara, 
	       CA, 93106-5050, USA}
\date{\today}

\begin{abstract}
Using first-principles density functional theory we 
predict a ferroelectric ground state -- driven by 
off-centering of the magnetic Mn$^{4+}$ ion -- in perovskite-structure 
BaMnO$_3$.
Our finding is surprising, since the competition between 
energy-lowering covalent bond formation, and energy-raising 
Coulombic repulsions usually only favors off-centering on the perovskite 
$B$-site for non-magnetic $d^0$ ions.
We explain this tendency for ferroelectric off-centering by 
analyzing the changes in electronic structure between the 
centrosymmetric and polar states, and by calculating the 
Born effective charges; we find anomalously large values 
for Mn and O consistent with our calculated polarization of 12.8 $\mu$C/cm$^2$. 
Finally, we suggest possible routes by which the  
perovskite phase may be stabilized over the usual hexagonal phase, 
to enable a practical realization of a single-phase multiferroic.

\end{abstract}

\pacs{71.15.Mb, 71.20.-b, 75.47.Lx, 77.80.-e}
\maketitle

Almost sixty years ago, Matthias observed that ferroelectricity occurs
in the $AB$O$_3$ perovskite structure when the $B$-site is a transition 
metal cation with a non-magnetic $d^0$ electronic 
structure\cite{Matthias:1949}.
Since then, many such $d^0$ perovskite ferroelectric compounds have been 
identified, and the requirement for ``$d^0$-ness'' has been 
explained in terms of covalent bond formation between empty transition 
metal $d$ and filled O 2$p$ orbitals\cite{Cohen:1992,Cohen/Krakauer:1992}.
This {\it Matthias rule} is problematic however for the design of new 
multiferroic perovskites, since magnetism is most readily accommodated 
in the perovskite structure by partial occupation of the transition metal 
$d$ orbitals. 
Indeed, the contra-indication between $B$-site ferroelectricity and 
$B$-site magnetism\cite{Hill:2000} has prompted the search for and 
identification of new mechanisms for ferroelectricity, which do not 
involve $B$-site cation off-centering: these include 
lone pair stereochemical activity\cite{Seshadri/Hill:2001}, spin 
spirals\cite{Newnham_et_al:1978}, charge ordering\cite{Ikeda_et_al:2005}, 
and geometric ferroelectricity\cite{vanAken_et_al:2004}.
In this work we discuss instead the circumstances under which the 
Matthias rule can be circumvented and magnetic $B$-site ions can be made 
to off-center, using perovskite-structure barium manganite as our example.
We begin this report with a review of the physics of the second order 
Jahn-Teller effect, which explains why a non-centrosymmetric 
distortion -- required for ferroelectricity -- is usually only favorable 
for $d^0$ transition metal cations. 
Then we discuss the balance between ferroelectric and competing 
non-ferroelectric distortions, which in perovskite oxides are often 
rotations of the oxygen octahedra. 
The main part of the paper is a detailed first-principles study of 
BaMnO$_3$ in the metastable perovskite structure: We find a 
ferroelectric ground state, with a substantial ferroelectric polarization, 
driven by off-centering of the magnetic Mn$^{4+}$ ion.
Finally, we discuss possibilities for realizing this structure experimentally.

\begin{figure}
\includegraphics[width=0.43\textwidth]{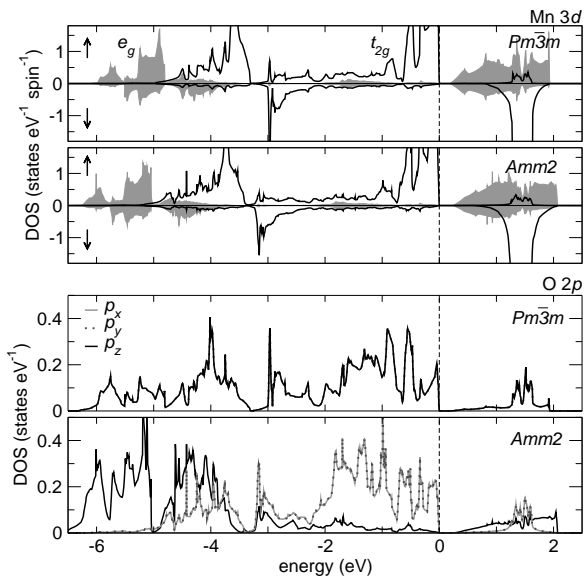}
\caption{\label{fig:dos} Orbital resolved Mn 3$d$ and O 2$p$
densities of states in the centrosymmetric cubic 
($Pm\bar{3}m$) and the polar ($Amm2$) structures. 
(Upper panels) spin-polarized DOS of a single Mn atom in the $G$-type 
AFM BaMnO$_3$ structure. (Lower panels) DOS of the O atom on top of 
the Mn atom.
In the $Pm\bar{3}m$ structure only the $p_z$ orbital is shown 
as the oxygen 2$p$ orbitals are degenerate.
}
\end{figure}

The tendency of a material to ferroelectric instability can be understood 
within the framework of vibronic coupling theory \cite{Bersuker:2001}, where 
it appears as the second-order terms in the perturbative expansion of the 
total energy with respect to distortions from a high symmetry reference phase. 
As a result it is often called the second-order Jahn-Teller (SOJT) effect.
Expanding the Hamiltonian as a function of normal coordinate $Q$ about the 
electronic Hamiltonian for the high symmetry reference phase, 
$\mathcal H^{(0)}$, gives
\begin{equation}
\mathcal H = \mathcal H^{(0)} + \mathcal H^{(1)}Q + 
\frac{1}{2}\mathcal H^{(2)}Q^2 + \ldots
\quad ,
\end{equation}
with
\[
\mathcal H^{(1)}  =  \left. \frac{\delta \mathcal H}{\delta Q} \right|_{Q=0}\: 
\mathrm{and} \; \; 
\mathcal H^{(2)}  =  \left. \frac{\delta^2 {\mathcal H}}{{\delta} Q^2}
\right|_{Q=0} \quad .
\]
$\mathcal H^{(1)}$ and $\mathcal H^{(2)}$ capture the vibronic coupling 
between the displacements of the ions from their positions in the high 
symmetry phase and the electrons.
Using standard perturbation theory, the energy can be expanded as a 
function of the normal coordinate about the high symmetry reference 
structure\cite{Burdett:1981,Pearson:1983} with energy $E^{(0)}$ as

\begin{eqnarray}
E & = & E^{(0)} + \bra{0} \mathcal H^{(1)} \ket{0}Q \nonumber \\
     & + & \frac{1}{2} 
[ \bra{0} \mathcal H^{(2)} \ket{0}
- 2 \sum_n \frac{| \bra{0} \mathcal H^{(1)} \ket{n} |^2}
              {E^{(n)} - E^{(0)}} ]Q^2 \nonumber \\
    & + & \ldots
\label{eqn:jt}
\end{eqnarray}
Here $\ket{0}$ is the the lowest energy solution of $\mathcal H^{(0)}$
and the $\ket{n}$s are excited states with energies $E^{(n)}$. 
The first-order term, $\bra{0} \mathcal H^{(1)} \ket{0}Q$, describes 
the regular first-order Jahn-Teller theorem. 
This term is non-zero only for orbitally degenerate states, and in the 
case of $d$ orbitals it always leads to centrosymmetric distortions, 
therefore it does not give rise to ferroelectricity. 
Note, however, that in cases where it is non-zero it dominates over any 
non-centrosymmetric second-order distortions. 
In non-orbitally degenerate systems, competition between the two second-order 
terms, which are of opposite sign, determines whether a non-centrosymmetric 
off-centering is favored or not.
The first of the two second-order terms describes the short-range 
repulsive forces which would result if the ions were displaced with 
the electrons frozen in their high symmetry configuration.
Since $\bra{0} \mathcal H^{(2)} \ket{0}$ is always positive, it always 
raises the energy of the system, and so polar distortions are more 
likely to be favored if this term is small; this tends to be the case 
for closed-shell $d^0$ cations without valence electrons.
The second of the second-order terms,  
$- \sum_n \frac{| \bra{0} \mathcal H^{(1)} \ket{n} |^2}
              {E^{(n)} - E^{(0)}} Q^2$,
describes the relaxation of the electronic system in response to the ionic 
displacements through covalent bond formation. 
It is always negative unless it is zero by symmetry, and so it favors 
ferroelectricity when its magnitude is large for non-centrosymmetric 
distortions. 
This occurs when the products of the ground and lowest excited states are 
of odd parity, so that the matrix elements are non-zero in the 
cases when $E^{(n)} - E^{(0)}$ is small.
For Mott insulators with partially filled $d$ shells, the top of the 
valence band and bottom of the conduction band are both composed primarily 
of transition metal $d$ states. 
Therefore the ground and low-lying excited states have the same symmetry, 
their product with $\mathcal H^{(1)}$ is odd, and the matrix element 
$\bra{0} \mathcal H^{(1)} \ket{n}$ is zero. 
Conversely, for $d^0$ perovskites, the top of the valence band is made 
up largely of O $2p$ states, and the bottom of the conduction band of 
transition metal $3d$ states, thus the product of the ground and low-lying 
excited states with $\mathcal H^{(1)}$
is even, and the matrix element $\bra{0} \mathcal H^{(1)} \ket{n}$  is non-zero.
Consequently, the balance between the positive and negative second-order 
terms usually results in off-centering for $d^0$ cations, such as Ti$^{4+}$ 
in the prototypical ferroelectric BaTiO$_3$. 
Here a strong increase in O 2$p$ -- Ti 3$d$ hybridization 
accompanies the distortion from the high symmetry to the 
polar structure\cite{Filippetti/Hill:2002} and so the relevant
$\bra{0} \mathcal H^{(1)} \ket{n}$ matrix elements are large.
This rearrangement of the electrons through covalent bond formation leads 
to Born effective charges, $Z^* = \delta P / \delta u$, which are 
significantly larger in magnitude than the formal charges on the ions; 
these are referred to as anomalous Born effective charges (BECs). 
An anomalous BEC is therefore a good indicator of the tendency of an ion 
to off-center and is often taken as a signature of ferroelectricity. 
The calculated BEC of Ti in the high-symmetry cubic phase of BaTiO$_3$, 
for example, is almost +7 whereas the formal 
charge is +4.\cite{Ghosez/Michenaud/Gonze:1998,Filippetti/Hill:2002}
This is because, as the Ti ion moves towards an oxygen carrying its 
positive charge, negative electronic charge flows towards it from the 
oxygen ion, leading to a larger polarization than would arise from 
the ionic component alone.
Conversely the balance between the second-order terms usually disfavors 
off-centering in transition metals with partially filled $d$ shells, since 
in this case the repulsive Coulomb interactions 
are stronger than any energy gain from chemical bond formation.
We note, however, that second-order Jahn-Teller behavior is referred to 
as an {\it effect}, not a theorem, because there is no fundamental 
requirement that the guidelines described above always hold.
(In contrast, first-order Jahn-Teller is a {\it theorem} since it is 
not a competition between two terms of opposite sign.)
In addition to this competition between bond formation and Coulomb 
repulsion that determines the tendency for off-centering distortions, 
other centrosymmetric instabilities compete with ferroelectric distortions 
in determining the ground state. 
We mentioned already first-order Jahn-Teller distortions which always 
dominate if they are allowed by symmetry. 
In addition, in perovskite-structure oxides centrosymmetric 
antiferrodistortive rotations of the oxygen octahedra are common. 
When these rotational instabilities are strong --usually in cases when 
the $A$-site ionic radius is smaller than ideal for the 
perovskite structure -- the tendency to form a polar ground state is drastically 
reduced even in materials with strongly anomalous $B$-site BECs. 
In fact, frustration of these rotational instabilities has been proposed 
as a viable route to novel ferroelectricity and 
multiferroic behavior.\cite{Bilc/Singh:2006,Singh/Park:2008} 
In this manuscript, we use first-principles calculations to identify 
a material -- perovskite structure BaMnO$_3$ -- in which a polar mode 
with transition-metal off-centering
is the dominant instability in spite of the partially filled $d^3$ manifold 
on the $B$-site. 
Our motivation for choosing perovskite BaMnO$_3$ is two-fold: 
First, CaMnO$_3$, which is paraelectric down to low temperature, 
has long been known to have an anomalous Mn BEC,\cite{Filippetti/Spaldin:2002} 
suggestive of a ferroelectric instability.
And second, recent calculations \cite{Bhatacharjee/Bousquet/Ghosez:2008,Bhattacharjee/Ghosez:2008} for CaMnO$_3$ showed that ferroelectricity can be artificially stabilized  
by increasing the lattice constants (equivalent to applying 
negative hydrostatic pressure) or with tensile strain, so that the short-range 
repulsions are substantially reduced. 
Since the Ba$^{2+}$ ion has a larger radius than Ca$^{2+}$, 
perovskite-structure BaMnO$_3$ is analogous to CaMnO$_3$ under negative 
pressure. 
We note, however, that because of the large size of Ba$^{2+}$, the experimental 
structure of BaMnO$_3$ is not perovskite, but rather a hexagonal phase with 
face-shared oxygen octahedra.\cite{Chamberland/Weiher:1970}
We discuss options for stabilizing the perovskite phase at the end of 
the manuscript.
Our first-principles density functional calculations are performed 
within the local spin-density approximation (LSDA) as implemented in the 
Vienna {\it ab initio} simulation package 
({\sc vasp}).\cite{Kresse/Furthmueller_PRB:1996,Kresse/Joubert:1999}
The projector augmented wave method\cite{Bloechl:1994} is used with  
the following valence electron configurations: 
5$s^2$5$p^6$6$s^2$ for Ba,  3$d^6$4$s^1$ for Mn and 2$s^2$2$p^4$ for oxygen.
The Brillouin zone integrations are performed with the tetrahedron 
method\cite{Bloechl/Jepsen/Andersen:1994} over a 
9$\times$9$\times$9  Monkhorst-Pack $k$-point mesh\cite{Monkhorst_Pack} 
centered at $\Gamma$ and a 450~eV plane-wave cutoff.
For structural relaxations a Gaussian broadening technique of 0.05~eV is 
used and the ions are relaxed until the Hellmann-Feynman forces are 
less than 1~meV~\AA$^{-1}$.
The electronic contribution to the polarization is calculated  
following the standard Berry phase 
formalism.\cite{King-Smith/Vanderbilt:1993,Vanderbilt/King-Smith:1993} 
We begin by determining the equilibrium volume for the hypothetical cubic 
(space group $Pm\bar3m$) BaMnO$_3$ structure with $G$-type antiferromagnetic 
order of the Mn$^{4+}$ cations.
This $G$-type antiferromagnetism is the likely ground state 
since both CaMnO$_3$ and SrMnO$_3$ have been shown to exhibit the same 
type of order in the perovskite phase.\cite{Wollan/Koehler:1955,Satpathy_et_al:1996}
We find that the primitive equilibrium unit cell volume $\Omega_0=58.7$~\AA$^3$ 
(cubic lattice parameter of $a_0 = 3.89$~\AA) is larger than 
previous first-principles calculations for CaMnO$_3$ 
using various exchange-correlation functionals, consistent with the 
larger $A$-site cation size.\cite{Pickett/Singh:1996,Hanfland_et_al:2006,
Bhatacharjee/Bousquet/Ghosez:2008}
The centrosymmetric structure is insulating with a LSDA band gap of 0.20~eV. 
(With a moderate effective Hubbard $U$ of 4.5~eV, we find a gap of 0.26~eV).
The valence band is predominantly majority spin Mn $t_{2g}$ and O 2$p$ 
character with strong $p$--$d$ hybridization 
[Figure \ref{fig:dos} (upper panels)].
The conduction band is formed by the empty Mn $e_g$ orbitals and the minority 
spin $t_{2g}$ states, which is consistent with Mn$^{4+}$ in an octahedral 
crystal field.
A local magnetic moment of 2.3~$\mu_B$ is found at each Mn site, which is 
slightly reduced from the formal 3~$\mu_B$ due to strong hybridization with the 
oxygen 2$p$ orbitals.
\begin{figure}[t]
\includegraphics[width=0.44\textwidth]{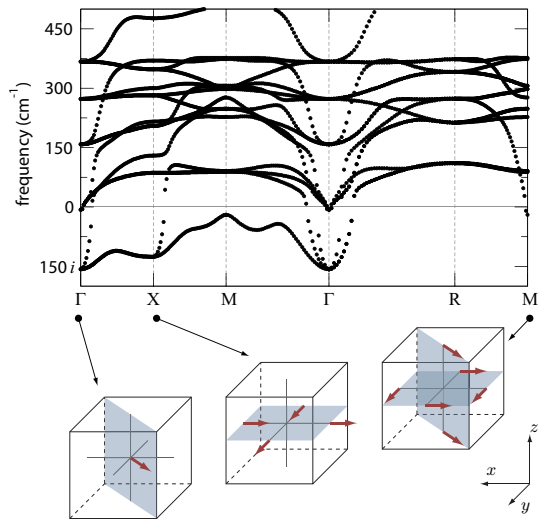}
\caption{\label{fig:phonon_disp} Phonon dispersions of the $G$-type 
AFM cubic BaMnO$_3$ along the high-symmetry lines. 
Energy lowering eigenvector modes for the unstable phonons at 
$\Gamma$, $\mathrm{X}$ and $\mathrm{M}$, with the Mn ion at the center of the unit cell 
surrounded by an oxygen octahedron.}
\end{figure}
We next calculate the lattice instabilities for the optimized cubic structure 
using the frozen-phonon method.\cite{Kunc/Martin:1982} 
In this method the phonons are 
found by calculating total energies with respect to atomic displacements from 
the reference structure at high symmetry positions, $q$, in the Brillouin zone.
We use a 3$\times$3$\times3$ (270 atom) supercell, which allows us to directly
access 8 high symmetry points in the Brillouin zone by freezing in different
atomic displacement patterns. For each $q$ value we construct the dynamical matrix 
from the Hellman-Feynman forces induced on the ions after making small positive and 
negative displacements (to remove any quadratic effects) about the 
high symmetry positions. 
Diagonalization of the dynamical matrix yields the atomic displacement patterns 
(eigenvectors) and phonon mode frequencies (eigenvalues) at that $q$ point.
The complete phonon dispersions are then determined by interpolating these
solutions of the dynamical matrix at the special $q$-points to the whole 
Brillouin zone using a Fourier 
interpolation scheme.\cite{Gonze/Lee:1997,fropho:2008}
The splitting between the longitudinal optic (LO) and transverse optic (TO) 
modes at $k=0$ is not included in our calculations.
In Figure \ref{fig:phonon_disp}, we show the phonon dispersion curves for 
$G$-type BaMnO$_3$ along the high-symmetry directions in the 
Brillouin zone of the primitive (5-atom) unit cell.
The dominant instability is a triply degenerate $\Gamma$-point mode with an 
imaginary frequency of 157.2$i$~cm$^{-1}$ and point symmetry $T_{1u}$. This
is a polar mode consisting of a relative Mn-O displacement in which the oxygen
octahedra remain almost rigid. In the ground state structure (described below) 
we find that this mode dominates with the Mn displacement occurring along the 
[110] direction of the primitive cubic unit cell.
Interestingly, the other high symmetry instabilities -- at the $\rm X$ and
$\rm M$ points -- do not correspond to the usual antiferrodistortive rotations
of the oxygen octahedra which are common in perovskites.
The $\rm X$-point mode has an imaginary frequency of 125.5$i$~cm$^{-1}$ and
$A_{2u}$ symmetry. It
corresponds to a breathing of the oxygen octahedra in the $xy$-plane in which
two adjacent oxygens move in towards the Mn ion and two move out, so that 
two short and two long Mn--O bonds are created (Figure \ref{fig:phonon_disp}). 
In the ground state structure
it combines with the $T_{1u}$ mode with almost equal amplitude.
The less unstable $\rm M$-point [$\omega=19.64i$, ($E_u$ symmetry)] 
also corresponds to a distortion rather than a rotation of the oxygen
octahedra, this time dominated by displacements of the apical oxygen atoms 
(Figure \ref{fig:phonon_disp}).  
%
We also verified that these modes are robust to correlation effects by 
repeating the calculations with an effective Hubbard $U$ parameter of 4.5~eV.
\begin{table}[b]
\begin{ruledtabular}
\begin{tabular}{llccc}%
Atom	&	Site	&	$x$	&	$y$	&	$z$  	\\
\hline
Ba	&	$2a$	&	0 &	0 &	$0.486$	\\
Mn	&	$2b$	&	$\frac{1}{2}$&	0	& $-0.018$  \\
O(1)	&	$2a$	&	0  &	0  &	$-0.011$	 \\
O(2)	&	$4e$	&	$\frac{1}{2}$&	$0.249$	& $-0.260$ \\
\end{tabular}
\end{ruledtabular}
\caption{\label{tab:lattice_info}Calculated structural parameters for BaMnO$_3$ 
with $Amm2$ symmetry. Our calculated lattice parameters are
are $a=3.84$, $b=5.43$, and $c=5.43$~{\AA} with a volume per
formula unit of 56.7~\AA$^3$. }
\end{table}
Next we determined the ground state structure by freezing linear 
combinations of each of the unstable modes described above into the
cubic $Pm\bar{3}m$ phase, then fully relaxing all internal degrees 
of freedom until the forces were less than 1~meV~\AA$^{-1}$.
We began with a constant volume constraint, and obtained a polar structure 
with $R3m$ space group.\cite{Stokes/isotropy}
We then lifted the volume constraint and performed a full optimization of 
the atomic positions and lattice parameters within orthorhombic symmetry 
(20-atom unit cell). 
We obtained a polar ground state structure with $Amm2$ space group, and 
structural parameters given in Table \ref{tab:lattice_info}.
The ground state structure is 64~meV lower in energy per formula unit than the cubic $Pm\bar{3}m$
structure. 
We find that our calculated ground state structure can be written as
a linear combination of the unstable modes of the cubic reference structure
with the following coefficients:
$0.694T_{1u}+0.604A_{2u}+0.393E_u$.
The resulting structure has three unique Mn--O bond lengths of 1.88, 1.92 
and 1.96~\AA\ with a mean bond length similar to the cubic case.
Furthermore, the mean Mn--O--Mn bond angle is decreased to 
178.2$^\circ$ from the ideal 180$^\circ$ cubic structure.
Next, we examine the changes in the electronic structure following displacement 
in order to explain the stabilization of the 
ferroelectric off-centering of the magnetic Mn cation.
In Figure \ref{fig:dos} we compare the densities of states (DOS) 
for the non-polar cubic and polar $Amm2$ structures; in particular, 
we focus on the orbital resolved DOS for the Mn 3$d$ and O 2$p$ bands.
The electronic structure of the metastable cubic phase was 
described earlier; here we reiterate that both the 
$t_{2g}$ and $e_{g}$ states are hybridized with the O 2$p$ 
orbitals indicating that some covalent bonding is already 
present in the cubic phase.
In the polar $Amm2$ ground state 
an increase in hybridization between the Mn $e_g$ and O 2$p$ levels 
occurs. This lowers the energy of the $e_g$ states by $\sim$0.30 eV, and 
shifts the O $p_z$ states from the top to the bottom of the valence band;
both factors result in an increase of the valence bandwidth.
%
%
%
%
%
In terms of our earlier discussion of the SOJT effect, the matrix elements
$\bra{0} \mathcal H^{(1)} \ket{n}$ are large because of the enhanced 
Mn $e_g$ and O 2$p$ hybridization, and the ferroelectric distortion is 
favored.
We next calculate the total polarization for the $Amm2$ structure 
as the sum of the ionic and electronic contributions 
(including both spin channels) using the Berry's phase method.
Here we find a value of 12.8~$\mu$C~cm$^{-2}$, which is substantial 
compared to many manganite multiferroics,\cite{Kimura_et_al_Nature:2003,Goto/Ramirez/Tokura:2004,Malashevich/Vanderbilt:2008} but consistent in magnitude 
with conventional $d^0$ ferroelectric perovskite oxides.
We next calculate the BECs for the cubic phase along the [111] direction, 
since large anomalies from the formal charge values are often 
good indicators for the underlying ferroelectric instability.
In Table \ref{tab:zstars} we tabulate the spin-resolved BECs for each atom $i$ 
decomposed such that $Z_i^{*} = Z_i^{\rm ion} + Z_i^{{\rm el},\uparrow} + 
Z_i^{{\rm el},\downarrow}$, where $Z_i^{\rm ion}$ is the pseudo-core charge.
\begin{table}[b]
\begin{ruledtabular}
\begin{tabular}{llllll}
 &	&	$Z^{*}$	& $Z^{\rm ion}$	&	$Z^{{\rm el},\uparrow}$	& $Z^{{\rm el},\downarrow}$ \\
\hline
Ba	&	Formal	& $+2$	& $10$	& $-4$	& $-4$ \\
		& LSDA		&	$2.72$	&	$10$	& $-3.64$& $-3.64$ \\
\hline
Mn	&	Formal	& $+4$	& $7$	& $-3$	& $0$ \\
  	& LSDA		&	$10.22$	&	$7$	& 	$+3.75$	& $-0.53$ \\
\hline
O 	&	Formal	& $-2$	& $6$	& $-4$	& $-4$ \\
  	& LSDA		&	$-3.97$	&	$6$	& 	$-4.99$	& $-4.98$ \\
\end{tabular}
\end{ruledtabular}
\caption{\label{tab:zstars} Spin-resolved Born effective charges for cubic BaMnO$_3$ 
calculated within the LSDA along the [111] direction. 
The spin components for Ba and O are equivalent within the 
numerical accuracy.
}
\end{table}
We find an anomalously large BEC for Mn, consistent with our finding
of a ferroelectric instability dominated by Mn displacement. In fact
we see that the majority spin electronic contribution is {\it positive},
corresponding to a net flow of electrons {\it towards} the Mn as it 
displaces towards the oxygen; this is consistent with the enhanced
Mn--O hybridization that we observed in our calculated DOSs. 
[An anomalously large value is also found for Mn (8.45), when correlation 
effects are added, albeit reduced due to the enhanced band gap 
and narrowing of the $e_g$ bandwidth.]
Interestingly, the minority spin electronic contribution on the Mn ion
is  close to the formal charge value; this reflects the low availability of
Mn minority states to accept electrons at the bottom of the conduction band 
in the cubic structure.
In the polar $Amm2$ structure, the calculated Mn $Z^*$ in the [111] Cartesian 
direction is 8.34 (7.45 within the LSDA$+U$ method); this reduction from the 
cubic case is consistent with behavior in conventional perovskites and also 
reflects in this case the opening of the band gap between the cubic and ground 
state structures.
We conclude by exploring the feasibility of accessing perovskite-structure 
BaMnO$_3$ experimentally.
As we discussed above, non-ferroelectric CaMnO$_3$ exists in the perovskite
structure and is nearly cubic\cite{Poeppelmeier_et_al:1982} with small rotations 
of ideal MnO$_6$ octahedra.
Increasing the size of the $A$-site cation, which we have shown promotes the 
tendency to ferroelectric instability, also destabilizes the perovskite structure. 
SrMnO$_3$ is found experimentally in both cubic perovskite and the non-perovskite 
2H hexagonal polymorphs\cite{Syono/Akimoto/Kohn:1969}, while BaMnO$_3$ is known
only in the denser hexagonal 4H structure with face-shared 
octahedra.\cite{Hardy:1962}
Therefore, we next calculate the relative stability of our predicted
$Amm2$ perovskite-structure BaMnO$_3$ and the experimental 4H structure. 
Our results are shown in Figure \ref{fig:vol_compare}, as a function of
pseudo-cubic lattice parameter.
We find that the experimental 4H-BaMnO$_3$ structure is more stable than 
the $Amm2$ perovskite structure by $\sim$0.50~eV per formula unit. 
Only near 4.7\% expansion of the 4H lattice parameter is the perovskite
structure energetically favored over the 4H structure.
\begin{figure}
\includegraphics[width=0.38\textwidth]{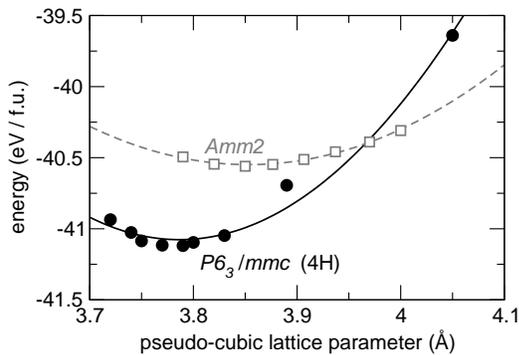}
\caption{\label{fig:vol_compare} Total energies for the  
the polar $Amm2$ and experimental 4H BaMnO$_3$ structures 
as a function of the pseudo-cubic lattice parameter.
The lines are quadratic fits to the calculated energies.
}
\end{figure}
These large energy differences suggest that experimental synthesis of the 
perovskite phase will be challenging, although might be feasible through
epitaxial stabilization in ultra-thin films; this approach has indeed been 
successful in stabilizing orthorhombic, over hexagonal, rare-earth
manganites \cite{Marti/Fontcuberta_et_al:2008,Salvador_et_al:1998}. 
Another possibility is to search for a critical alloying range
with Ca and/or Sr at which the perovskite structure remains stable
over the 2H or 4H structure, but within the perovskite structure where 
the ferroelectric instability has already started to dominate. 
Finally, we note that, if a ferroelectric Ba-based perovskite manganite
could be achieved experimentally, subsequent rare-earth doping on the
$A$-site could lead to an intriguing analogy to the colossal magnetoresistive 
(CMR) materials. In conventional CMR materials, a phase transition between a 
paramagnetic insulator and a ferromagnetic metal is induced by applied
magnetic field and/or temperature. In rare-earth doped BaMnO$_3$, the
insulating phase might be expected to be polar, leading to a
ferroelectric insulator -- ferromagnetic metal transition. We suggest
Ba-rich La$_{1-x}$Ba${_x}$MnO$_3$ as a material for further
experimental investigation.
To summarize, we have used hypothetical perovskite-structure BaMnO$_3$ to 
demonstrate that, in contrast to conventional wisdom, non-$d^0$ magnetic 
cations can undergo SOJT off-centering distortions resulting in polar 
ground states.
We have explained the behavior in terms of the delicate balance between 
competing energies in the second-order Jahn-Teller effect, and shown that
reducing Coulombic repulsions via increasing ionic separations is a general
route to promoting ferroelectricity in magnetic compounds. 
Finally, we have explored the feasibility of realizing such novel single-phase 
multiferroics experimentally.


The authors thank A.J.\ Hatt and S.\ Bhatacharjee for useful discussions.
This work was supported NSF under grant no.\ DMR-0605852 (NAS).
JMR acknowledges support from NDSEG (DoD), the IMI Program of the 
National Science Foundation under award no.\ DMR04-09848, 
and hospitality from the Department of Advanced Materials, University of 
Tokyo, Kashiwanoha. 
ASE was supported by the Apprentice Researchers program of the CNSI 
at UCSB with funding from the NNIN (award no.\ 44771-7475).
Portions of this work made use of the SGI Altix {\sc cobalt} system 
at the National Center for Supercomputing Applications under grant 
no.\ TG-DMR-050002S.


\end{document}